# Stepwise Quenching of Exciton Fluorescence in Carbon Nanotubes by Single Molecule Reactions

Laurent Cognet[1,2*], Dmitri A. Tsyboulski[2†], John-David R. Rocha[2†], Condell D. Doyle[2], James M. Tour[2] and R. Bruce Weisman[2*]

[1]*Centre de Physique Moléculaire Optique et Hertzienne, Université Bordeaux 1, and CNRS, 351 cours de la libération, Talence, F-33405 France*

[2]*Department of Chemistry, Center for Biological and Environmental Nanotechnology, and R. E. Smalley Institute for Nanoscale Science and Technology, Rice University, 6100 Main Street, Houston, Texas 77005*

The chemical reactions of single molecules with individual single-walled carbon nanotubes are observed, and luminescence quenching analysis reveals the diffusional range of excitons in semiconducting nanotubes.

*To whom correspondence should be addressed: lcognet@u-bordeaux1.fr (LC), weisman@rice.edu (RBW)

[†]These authors contributed equally to this work.

**Abstract**

Single-molecule chemical reactions with individual single-walled carbon nanotubes were observed through near-infrared photoluminescence microscopy. The emission intensity within distinct submicrometer segments of single nanotubes changes in discrete steps after exposure to acid, base, or diazonium reactants. The steps are uncorrelated in space and time, and reflect the quenching of mobile excitons at localized sites of reversible or irreversible chemical attack. Analysis of step amplitudes reveals an exciton diffusional range of about 90 nanometers, independent of nanotube structure. Each exciton visits approximately $10^4$ atomic sites during its lifetime, providing highly efficient sensing of local chemical and physical perturbations.

Optical excitation of semiconducting single-walled carbon nanotubes (SWNTs) generates relatively strongly bound excitons that have spatial dimensions predicted to be a few nanometers (*1-3*). Experimental evidence of efficient exciton-exciton annihilation in nanotubes indicates that SWNT excitons have significant mobility along the tube axis (*4-6*). However their mobility is still experimentally and theoretically uncertain. A notable related effect is the strong suppression of photoluminescence (PL) when SWNT sidewalls are perturbed by chemical reactions (*7-9*). This quenching phenomenon has hampered the use of covalently derivatized SWNTs as near-infrared (near-IR) fluorophores.

We report here the use of single-nanotube microscopy to detect stepwise changes in SWNT PL intensity within segments of individual nanotubes while they are exposed to chemical reactants. These stepwise changes in PL intensity are caused by reactions of single molecules with one nanotube. Since the pioneering low-temperature experiments of Orrit and Moerner, single-molecule spectroscopy has proven to be a powerful tool that bypasses ensemble averaging in the study of static and dynamic nano-objects in various environments (*10,11*). Single-molecule approaches are especially appealing for SWNT fundamental studies and applications (*12-14*) because the bulk samples are highly heterogeneous. In the present work, the magnitudes of PL intensity steps caused by single molecule reactions reveals that the exciton excursion range in highly luminescent SWNTs is ~ 90 nm and essentially independent of nanotube structure. This room-temperature excitonic motion is deduced to be diffusional. Because each nanotube exciton visits a very large number of atomic sites during its lifetime, PL quenching provides an ultrasensitive method for sensing and studying certain types of chemical reactions with nanotube sidewalls at the single-molecule level.

Our studies required highly luminescent and relatively long nanotubes that were immobilized yet accessible to added reactant solutions. We therefore used very brief tip ultrasonication to disperse raw HiPco SWNTs in aqueous sodium dodecylbenzenesulfonate (SDBS) surfactant before mixing the dilute suspension with a melted agarose gel preparation (*15*). Aqueous gels are commonly used in molecular biology to provide an inert immobilizing environment and they have previously been applied in single-molecule optical measurements (*16*). We captured near-IR fluorescence images of single nanotubes with a wide-field inverted microscope modified to include laser excitation and a low-noise InGaAs camera (*14*). Micrometers-long individual SWNTs were easily identified in these images and some of them displayed uniform and high PL intensity (Fig. 1A). The (*n,m*) structural assignment of each studied tube was deduced from the peak wavelength of its narrow Lorentzian emission spectrum (*17*), as measured with a dedicated spectrograph and multichannel near-IR detector (Fig. 1B). In this study, we selected the brightest nanotubes present in the weakly sonicated samples. We believe that these selected nanotubes have low defect densities and nonradiative relaxation rates that are dominated by intrinsic processes. Most of these nanotubes displayed PL that varied linearly at the excitation intensities used here and remained stable for tens of minutes under continuous illumination (Fig. 1B and C). We used low laser intensities (~100 W/cm$^2$) to avoid exciton-exciton annihilation effects that would shorten the exciton lifetime and reduce the exciton's excursion range (*4-7*). Furthermore, we verified that the nanotubes studied here showed no detectable spectral shifts, spectral broadening, or change in emission intensity during the typical duration of our experiments (Fig. 1B). Finally, we confirmed that the emission spectra of SWNTs embedded in agarose gels matched those of SWNTs in fluid surfactant suspension.

To initiate chemical reactions, we placed a 10-µL drop of reactant solution on the edge of the 20 mm by 20 mm sample gel while recording fluorescence images of a specific nanotube with stable and spatially uniform emission (Fig. S1). Two reactants were used. One was sulfuric acid (0.05 M), which causes pH-dependent, reversible quenching of SWNT fluorescence (*7,8,18*). The other was 4-chlorobenzenediazonium tetrafluoroborate (1 mg/mL in water), which is known to give irreversible covalent derivatization of carbon atoms on the nanotube sidewall (*19*). We recorded fluorescence image sequences at an 18 Hz frame rate and plotted the total intensity within selected 2 × 2 pixel regions to obtain PL intensity traces. The chosen spatial binning region (4 pixels) corresponded to 670 nm by 670 nm, which is essentially the diffraction-limit at the near-IR wavelengths.

The PL intensity trace from a 670 nm segment of an individual (8,6) nanotube after exposure to sulfuric acid is shown on Fig. 2A. That nanotube's luminescence spectrum is plotted in Fig. 2B. Coarsely, the PL time trace displays an exponential decay form. However, closer examination reveals a series of strikingly distinct steps between well-defined intensity levels, rather than a continuous decrease (see the inset of Fig. 2A). Moreover, the luminescence shows both upward as well as downward steps. In order to analyze the distribution of step amplitudes, we compiled a histogram for the first 108 s period showing signal differences ($\delta I$) between successive image frames (Fig. 2C). Apart from the large main peak near zero, which arises mainly from signal noise, there are four distinct side bands distributed symmetrically around zero. These side bands are the signature of specific step amplitudes. The inner two, (labeled +1 and -1) correspond to positive and negative steps of the same unit intensity, whereas the outer ones (+2 and -2) correspond to double-amplitude steps. The presence of discrete (quantized) intensity changes is confirmed by the precisely linear relation linking the amplitudes of these intensity steps (Fig. 2D).

We attribute the observed discrete intensity steps to individual protonation reactions at the nanotube surface. The -1 and -2 steps would arise, respectively, from one or two independent sidewall reactions occurring within the 54 ms sampling interval, whereas the +1 and +2 steps arise from the reverse chemical processes. The mechanism through which acids quench nanotube luminescence has been described as a protonation at the sidewall of the nanotubes (*7,8,18*). In this scheme, a hole is injected into the nanotube $\pi$-system near the protonation site. If an exciton encounters such a chemically induced hole before it radiatively recombines, the exciton's luminescence will be efficiently quenched through nonradiative Auger processes (*6*).

Because the excitons have a limited excursion range, only a fraction of those generated within an observed 670 nm-long segment will be quenched by a single protonation-induced hole. If the density of such holes is low, then each one causes the same amount of quenching (corresponding to the observed -1 step amplitude), and each de-protonation restores the same amount of luminescence (the observed +1 step amplitude). However, when the typical distance between quenching sites becomes shorter or comparable to the exciton excursion range, the quenching impact of additional sites is reduced, giving smaller and eventually unresolved steps, as seen at later times in Fig. 2A.

To confirm that the observed stochastic stepwise quenching arises from reversible protonation, we have added ~10 μL of 1 M NaOH to nanotube samples that had previously been quenched by acid addition. The down triangle in Fig. 2E marks the point of acid addition, and the up triangle marks that of base addition. The period of luminescence restoration is expanded in Fig. 2F, which shows the expected stochastic stepwise intensity changes. Statistical analysis of upward and downward steps in a medium of known pH can in principle provide the pKa of the protonated nanotube. However, we could not perform this analysis on our data because of time-dependent pH gradients in the sample. Large intensity fluctuations are observed after luminescence has been restored by base addition. We attribute this behavior to local pH fluctuations near the nanotube within the agarose pores after the successive addition of acid and base without active mixing. Finally, the higher intensities seen in Fig. 2E after restoration likely reflect a higher final pH level at the nanotube (*18*).

To compare these exciton quenching effects with those caused by an irreversible chemical reaction, we performed similar measurements on samples exposed to diazonium salts (Fig. 3) (*20*). These traces also display distinct steps in luminescence intensity, but the great majority of the steps are negative (>95%), indicating a mainly irreversible process (*9*). Interestingly, adjacent diffraction-limited segments within a single nanotube exhibit very different dynamics, with little or no correlation between the time positions of the steps (compare each set of red and black traces in Fig. 3, A to C). Similar spatially uncorrelated behavior in acid quenching is vividly illustrated by the video in Supporting Online Materials (*15*). This observation demonstrates that the mean excursion range of excitons must be smaller than our optical resolution of 670 nm. The same conclusion can be drawn from the localized PL in bent nanotubes excited by linearly polarized light ((*14*) and Fig. S2).

Comparison of the PL time traces within each frame of Fig. 3, A through C reveals that for different segments of one nanotube, the first few step heights are integer multiples of the same unit intensity change $\Delta$ (regions marked by a single star). This stepwise quenching pattern leads to emission intensity plateaus that match between distinct segments of the same tube. As was observed for acid quenching, the steps have identical amplitudes at low densities of quenching sites, and gradually smaller amplitudes at longer times when the typical distance between quenching sites becomes shorter than or comparable to the exciton excursion range. However, in contrast to the reversible protonation reaction of acid quenching, the single-molecule diazonium salt reactions are essentially irreversible. Here, the chlorobenzenediazonium ion (or its diazotate dimeric product) is thought to first adsorb onto the surface of a nanotube, forming a charge transfer complex. The complex then irreversibly decomposes to form a covalent bond with the nanotube surface (*9,21*).

We presume that either of these nanotube perturbations can cause efficient exciton nonradiative decay. A mechanism for luminescence quenching by sidewall derivatization is suggested by scanning tunneling spectroscopy results on nanotubes covalently functionalized with amido groups, obtained with 10 nm resolution. The semiconducting energy gap was locally filled at the derivatization site (*22*). Excitons encountering such electronic perturbations should undergo efficient nonradiative recombination.

For each nanotube studied, repeatable values of $\Delta$ are obtained at early stages of the reaction when only a few independent sites have been derivatized within the observed diffraction-limited segment. Significantly, we find that $\Delta$ values match for separate segments in a single nanotube (Fig. 3, A to C). As for the minor variations in intensity step heights found at early stages of the reaction, we attribute these to the finite length of the diffraction-limited segment, since quenching events near the edge of the region produce smaller steps. Interestingly, several simultaneous steps sometimes occur within a single 54 ms frame, (Fig. 3 A and B). This behavior probably results from nonuniform dilution of reactants in the sub-micrometer pores of the agarose gels, exposing the nanotube segments to large local concentration fluctuations.

Normalizing $\Delta$ by the initial luminescence intensity, $I$, gives a ratio representing the probability that a mobile exciton within the nanotube segment encounters a chemical reaction site during its lifetime. This probability also represents the fraction of the nanotube length explored by an exciton. The average exciton excursion range, $\Lambda$, is therefore simply given by $(\Delta/I) \times L$, where $L$ is the observed segment length. For $L = 670$ nm, we have measured $\Lambda$ values for 19 different highly luminescent SWNTs. The resulting distribution is strikingly narrow, with a mean value of 91 nm (Fig. 4A). The small dispersion found for $\Lambda$ values suggests that $\Lambda$ does not depend strongly on nanotube structure. We plotted $\Lambda$ values measured for 11 different ($n,m$) species (*23*) as a function of SWNT diameter (Fig. 4B) and found no systematic variation of exciton excursion range with diameter (for the 0.7 to 1.1 nm range studied) or with chiral angle.

As mentioned above, our determination of Λ assumes that the local perturbation induced by a reaction sufficiently modifies the electronic structure of these bright tubes (*22*) to completely quench excitons at that site. If this quenching probability, *P*, is in fact less than 1, then Λ would be given by 90 nm / P (but could not exceed 670 nm). However, several observations support the estimate of *P* = 1. The results are independent of nanotube structure; the normalized initial step amplitudes are consistent for acid quenching and diazonium salt quenching; and Auger quenching of excitons is known to be efficient,

Is the exciton motion is ballistic or diffusional? For the bright emissive SWNTs studied here, we can estimate the exciton recombination lifetime, $\tau$, as 100 ps, which is among the larger values reported from time-resolved spectroscopy (*24,25*), If the motion is ballistic, then the exciton's kinetic energy would be given by $E_{ballistic} = \frac{1}{2} m^*_{exciton} \left(\frac{\Lambda}{\tau}\right)^2$, where $m^*_{exciton}$ is the exciton effective mass. Using an estimated $m^*_{exciton}$ of ~0.1 $m_e$ (*1,26*), we obtain $E_{ballistic}$ ~ $10^{-7}$ eV. This value is orders of magnitude lower than the 0.012 eV (½ $k_B$T) expected from rapid thermalization induced by the exciton-phonon coupling evident in SWNT spectroscopy (*27*). We conclude that the excitonic motion is instead diffusional, in agreement with a recent investigation on bulk samples (*28*). The corresponding diffusion coefficient for the exciton 1-D random walk is then given as $D = \Lambda^2/2\tau$ ~ 0.4 cm$^2$ s$^{-1}$.

This value lies more than two orders of magnitude below a prior tentative estimate that was indirectly deduced from transient absorption experiments on ensembles of short SWNTs in annealed dried films (*29*). The exciton diffusion coefficient measured here for highly luminescent individual SWNTs may be used to interpret the ~150 cm$^{-1}$ Lorentzian linewidths observed in emission spectroscopy of room-temperature SWNTs (see Figs. 1B and 2B). These widths arise from dephasing that is far more rapid (70 fs) than the ~100 ps exciton population lifetime $\tau$.

If dephasing instead occurs by diffusional hopping of the exciton along the tube axis, then the hopping step length $d_{hop}$ can be estimated from the relation $D = d_{hop}^2 / 2\tau_{hop}$. Here, $\tau_{hop}$ represents the interval between hops, identified as the 70 fs dephasing time. Interestingly, this approach gives a hopping step length of ~2 nm, which closely matches theoretical estimates of the exciton size (*1-3*). We propose that the emission linewidth may originate from thermally assisted, short-range hopping of excitons along the nanotube axis. At cryogenic temperatures, suppression of this hopping would then remove the major source of line broadening and account for dramatically narrower spectral features (*30*).

If efficient exciton quenching occurs not only at chemical derivatization sites, but also at the ends of cut nanotubes, our value of the exciton diffusion range, $\Lambda$, naturally explains the strong reduction in PL reported for SWNTs shorter than ca. 100 nm (*31*). More generally, the measured diffusion range implies that one exciton will visit ca. $10^4$ carbon atoms during its lifetime. Thus PL becomes remarkably sensitive to certain sidewall electronic perturbations, which may include chemical derivatizations and defects introduced during nanotube processing. PL may then prove useful for detecting local pH gradients in restricted environments, such as microfluidic channels or organelles inside biological cells.

FIGURE CAPTIONS

**Fig. 1.** Fluorescence measurements on an individual SWNT immobilized in agarose gel. (**A**) Image of an (11,3) SWNT excited at 100 W/cm². Scale bar is 1μm, and the intensity scale is linear from 0 (black) to 45,000 camera counts (white). (**B**) Spectrum of the same nanotube (top), and color-coded contour plot (bottom) showing the stability of spectra taken at 5 s intervals for 300 s. (**C**) Luminescence intensity from the same nanotube as a function of excitation intensity. Linearity up to 500 W/cm² indicates that no multi-exciton processes occur at 100 W/cm², which is the excitation intensity used in this study.

**Fig. 2.** Reversible stepwise quenching of individual SWNTs by acid. (**A**) Luminescence intensity of a diffraction-limited segment of an individual (8,6) SWNT following addition of acid. Inset shows an expansion of the first 100 s period, revealing well-defined reversible steps. (**B**) Near-IR emission spectrum of this SWNT, showing a single Lorentzian peak. (**C**) Histogram constructed from the data in the inset of (**A**), showing the distribution of changes in luminescence signal ($\delta I$) between successive 50 ms frames. Four symmetrically located sidebands are evident, at well defined $\delta I$ values. (**D**) Sideband peak positions as a function of sideband label. The linear relationship indicates that the sidebands correspond to discrete protonation and de-protonation steps (see text). (**E**) Luminescence signal of a diffraction-limited segment of an individual (7,6) SWNT upon successive addition of acid (▼) and base (▲). Luminescence recovery is observed after base addition. (**F**) Expanded inset from (**E**) showing reversible steps in the recovery period.

**Fig. 3.** Irreversible localized quenching of individual SWNTs with 4-chlorobenzenediazonium tetrafluoroborate**.** Luminescence quenching of diffraction-limited segments of individual SWNTs of type **(A)** (8,3), **(B)** (8,6) and **(C)** (11,1) upon addition of the diazonium salt solution. Irreversible intensity steps are observed. Each star (*) marks an elementary unit step. In each frame, the red (lower) and black (upper) traces correspond to adjacent segments (670 nm length each) of the same nanotube. The black traces have been shifted upward by one unit step for clarity.

**Fig. 4.** Exciton excursion range ($\Lambda$) measured for various SWNTs. **(A)** The distribution (n = 19) of $\Lambda$ values deduced from elementary step amplitudes (see text). **(B)** Plot of $\Lambda$ values as a function of SWNT diameter.

References and Notes

1. T. G. Pedersen, *Phys. Rev. B* **67**, 073401-073401 /4 (2003).

2. V. Perebeinos, J. Tersoff, Ph. Avouris, *Phys. Rev. Lett.* **92**, 257402-257402/4 (2004).

3. C. D. Spataru et al., *Phys. Rev. Lett.* **92**, 077402-077402/4 (2004).

4. J. Chen et al., *Science* **310**, 1171 (2005).

5. Y.-Z. Ma et al., *Phys. Rev. Lett.* **94**, 157402-157402/4 (2005).

6. F. Wang et al., *Phys. Rev. B* **70**, 1 (2004).

7. M. O'Connell et al., *Science* **297**, 593 (2002).

8. M. S. Strano et al., *J. Phys. Chem. B* **107**, 6979 (2003).

9. M. L. Usrey, E. S. Lippmann, M. S. Strano, *J. Am. Chem. Soc.* **127**, 16129 (2005).

10. W. E. Moerner and M. Orrit, *Science* **283**, 1670 (1999).

11. L. Cognet et al., *Science's STKE*, http://stke.sciencemag.org/cgi/content/full/sigtrans;2006/327/pe13 (21 March 2006).

12. A. Hartschuh et al., *Science* **301**, 1354 (2003).

13. J. Lefebvre et al., *Phys. Rev. B* **69**, 075403-075403/5 (2004).

14. D. A. Tsyboulski, S. M. Bachilo, R. B. Weisman, *Nano Lett.* **5**, 975 (2005).

15. Materials and methods are available as supporting material on Science Online.

16. R. M. Dickson et al., *Science* **274**, 966 (1996).

Figure 1

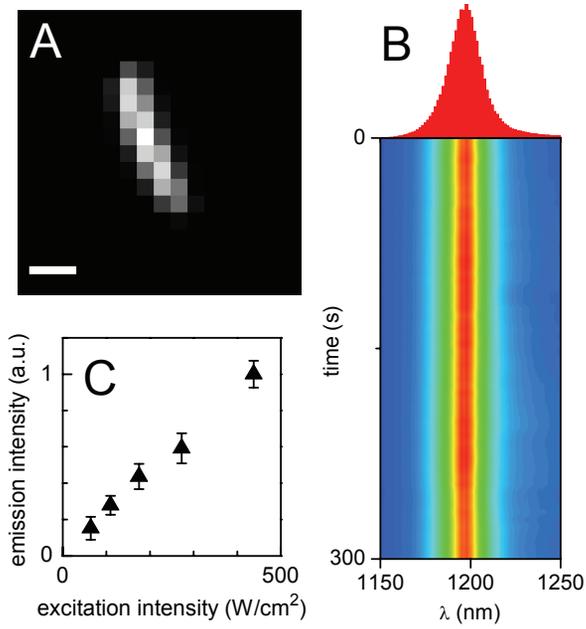

Figure 2

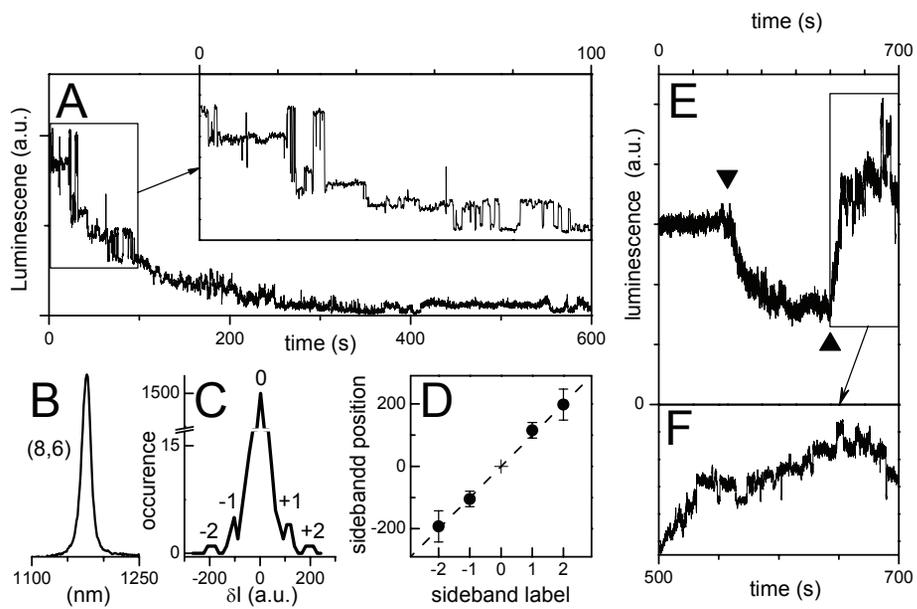

Figure 3

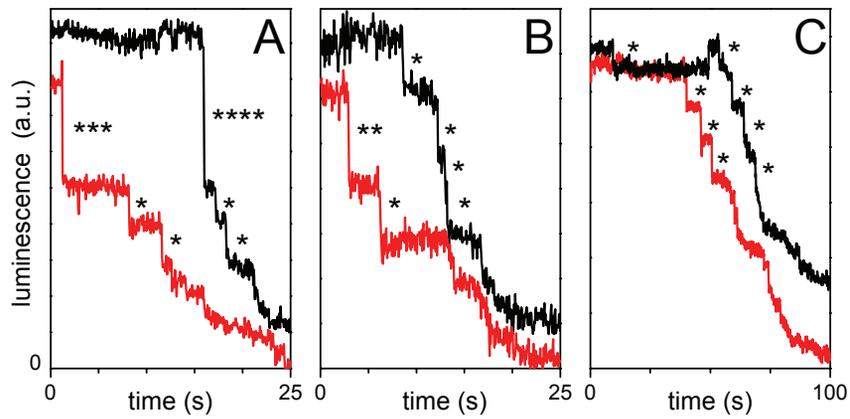

Figure 4

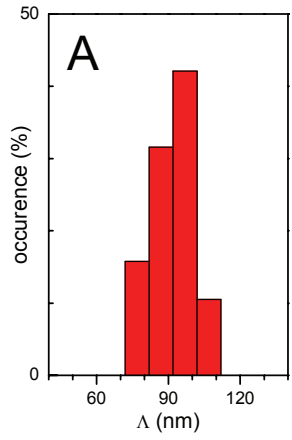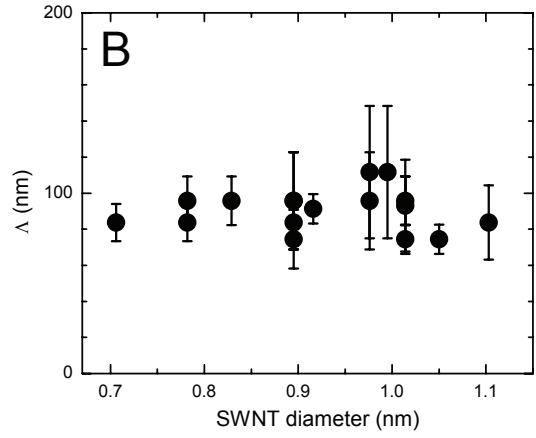